\documentclass[journal]{rmaa}

\def\uv{uvby-\beta \ }
\def\c1{c_1}

\begin{document}

\title{Physical parameters of seven field RR Lyrae Stars in Bootes\altaffilmark{1}}

\author{J. H. Pe\~ na\altaffilmark{2}, A. Arellano Ferro\altaffilmark{2}, R. Pe\~na Miller\altaffilmark{3}, J. P. Sareyan\altaffilmark{4} and M. Alvarez\altaffilmark{5}}
\medskip

\altaffiltext{1}{Based on observations collected at the San Pedro
M\'artir Observatory, Mexico} \altaffiltext{2}{Instituto de
Astronom\'{\i}a, Universidad Nacional Aut\'onoma de M\'exico.}
\altaffiltext{3}{Department of Mathematics, Imperial College London,
UK.}\altaffiltext{4}{Laboratoire Gemini, Observatoire de la Cote
d'Azur, France.}\altaffiltext{5}{Observatorio Astron\'omico
Nacional, Instituto de Astronom\'{\i}a, Universidad Nacional
Aut\'onoma de M\'exico}

\fulladdresses{
\item J.H. Pe\~na, A. Arellano Ferro:
Instituto de Astronom\'{\i}a, Universidad Nacional Aut\'onoma de
M\'exico, Apdo. Postal 70-264, M\'exico D. F. CP 04510, M\'exico.
\item M. Alvarez: Instituto de Astronom\'{\i}a, Universidad
Nacional Aut\'onoma de M\'exico, Apto.Postal. 877, Ensenada, BC
22860, M\'exico.
\item J. P. Sareyan: OCA, Department Gemini, Boulevard de
l'Observatoire BP 4229, 06304 Nice Cedex 4, France.
\item R. Pe\~na Miller: Department of Mathematics, Imperial College London,
South Kensington Campus, London SW7 2AZ, UK}

\shortauthor{Pe\~na et al.} \shorttitle{Seven RR Lyrae stars in
Bootes}

\resumen{Se reporta fotometr\'ia $uvby-\beta$ para
las estrellas tipo RR Lyrae AE, RS, ST, TV, TW, UU, y XX en Bootes. Se
calculan los par\'ametros f\'isicos $M/M_{\odot}$,
$\log(L/L_\odot)$, $M_V$, $\log~T_{\rm eff}$ y [Fe/H] a partir de
la descomposici\'on de Fourier de las curvas de luz y de
calibraciones emp\'iricas desarrolladas para este tipo de estrellas,
y se discute la confiabilidad de
estos valores. Los valores de [Fe/H] obtenidos se comparan con aquellos
calculados a partir del \1ndice $\Delta S$ para algunas estrellas de la muestra.
Se encontr\'o
que el enrojecimiento de la zona es despreciable, compar\'andolo
con el mostrado por diversos objetos en la misma regi\'on del
cielo. Por tanto, se calcularon las distancias a estos objetos.
La variaci\'on, a lo largo del ciclo de pulsaci\'on, de los \1ndices fotom\'etricos
desenrojecidos $(b-y)_o$ y $c_1$ permite la comparaci\'on con mallas
te\'oricas y por lo tanto la estimaci\'on independiente de  $\log~T_{\rm eff}$ y
${\rm log}~g$.}

\abstract{Str\"omgren $uvby-\beta$ photometry is reported for the RR
Lyrae stars AE, RS, ST, TV, TW, UU, and XX in Bootes. The physical
parameters $M/M_{\odot}$, $\log(L/L_\odot)$, $M_V$, $\log~T_{\rm
eff}$ and [Fe/H], have been estimated from the Fourier
decomposition of the light curves and the empirical calibrations
developed for this type of stars. Detailed behavior of the stars
along the cycle of pulsation has been determined from the observed
photometric indices and the synthetic indices from atmospheric
models. The reddening of the zone is found to be negligible, as
estimated from the reddening of several objects in the same region
of the sky. Hence the distances to the individual objects are also
estimated. }

\keywords{Variable Stars-RR Lyrae, photometry, $\uv$ photometry }

\maketitle

\section{Introduction}
\label{sec:intro}

It is known that the Horizontal Branch (HB) morphology in globular clusters of similar metallicities varies and the identification of a parameter, other than metallicity, responsible for this has originated much discussion in the astronomical literature. Recent findings that more massive clusters tend to
have HBs extended further into higher temperatures led Recio-Blanco et al. (2006) to suggest the cluster's total mass may be  the "second parameter". However since the HB morphology varies
 due to stellar evolution, age is still a strong candidate for the
 second parameter status
(e.g. Stetson et al. 1999; Catelan 2000 and references therein).

RR Lyrae stars are outstanding stars in the HB whose absolute magnitudes and iron abundances have been
carefully calibrated (Clementini et al. 2003; Layden 1994) and therefore play an important role in studies of the structure of the Galaxy.
The estimation of individual physical parameters of RR Lyrae stars in globular clusters provides relevant insights not only into the distance and iron abundances of their parent clusters (e.g. Arellano Ferro et al. 2008a; 2008b), but also imposes constraints on the structure of the HB and on the stellar evolution in this phase (see e.g. Zinn, 1993; van den Bergh 1993; Zinn 1996).

Determination of the physical parameters in RR Lyrae stars can
be attained from the analysis of data obtained
from fundamental techniques in astronomy such as photometry or
spectroscopy. High dispersion spectroscopy produces accurate estimates
for some of the atmospheric physical parameters, i.e.,
temperature, surface gravity and metallicity, that have been used to calibrate
low resolution methods such as the $\Delta S$ method (Preston 1959). However this approach is
limited mostly to brighter objects since the extended telescope time required
to study fainter objects is very
competitive and hence scarce. However,
isolated examples of accurate spectroscopy of objects around  $V
\sim 16-17$~mag do exist (e.g. James et al. 2004, Gratton et al. 2005,
Yong et al. 2005, Cohen et al. 2005).

Alternative approaches to physical parameter estimations are
theoretical and semi-empirical calibrations of photometric indices,
such as the synthetic color grids from atmospheric models (e.g.
Lester, Gray \& Kurucz 1986) or, for RR Lyrae stars,
the decomposition of light curves in Fourier harmonics and the
calibration of the Fourier parameters in terms of key physical
parameters (e.g. Simon \& Clement 1993; Kov\'acs  1998; Jurcsik 1998,
Morgan et al. 2007). Kov\'acs and his collaborators, in an extensive
series of papers (e.g. Kov\'acs \& Jurcsik 1996, 1997;
Jurcsik \& Kov\'acs 1996; Kov\'acs \& Walker, 2001), have
developed purely empirical relations from the Fourier analysis of
the light curves. The basic stellar parameters are based on the
assumption that the period and the shape of the light curve are
directly correlated with the physical parameters (or quantities
related to them) such as the iron abundance [Fe/H], absolute magnitude
$M_V$ and effective temperature $T_{\rm eff}$.

In this paper we shall use the Str\"omgren photometry of seven RR
Lyrae stars in Bootes to estimate their physical parameters.
Furthermore, since we have obtained simultaneous $uvby-\beta$
photoelectric photometry, the precise variation of the stellar
brightness and colors along the cycle of pulsation for
each star can also be used to provide rigid constraints
on the physical parameters.

\section{Observational material and Reductions}
\label{sec:Observations}

Some of the stars included in the present work were part of
the observing program in previous studies as secondary targets
(Lampens et al. 1990 and Pe\~na 2003). Other seasons were devoted
entirely to some of these RR Lyrae stars. The log of the various seasons
is given in Table \ref{LOG}. The observations were obtained with the 1.5 m
telescope at the Observatorio Astron\'omico Nacional of San Pedro
M\'artir, Mexico. The telescope was equipped with a six channel
spectrophotometer described in detail by Schuster \& Nissen (1988).
All the data reductions were performed using the NABAHOT
package (Arellano Ferro \& Parrao 1989).

\begin{table*}[!ht]
\small{
\begin{center}
\caption[] {\small Log of the observing seasons} \hspace{0.01cm}
    \label{LOG}
\begin{tabular}{lcc}
\hline \hline
Star  &  Initial date  &  Final date \\
\hline  \hline
AE Boo   &  Jun 08, 2001 & Jun 11, 2001 \\
TV Boo   &  Feb 04, 2002 & Feb 11, 2002 \\
RS Boo \& TV Boo &  Apr 08, 2004 & Apr 11, 2004 \\
Full sample&  May 27, 2005&Jun 16, 2005\\
\hline \hline
\end{tabular}
\end{center}
}
\end{table*}

In those seasons in which the present RR Lyrae stars were supplementary
objects, fewer data
points were obtained each night.
The accumulated time span including all the seasons, however, was
large enough to cover the whole cycle of all the stars in our sample.
In each season the same
observing routine was employed: a multiple series of integrations
was carried out, often five 10 s integrations of the star, to the
average of which a one 10 s integration of the sky was subtracted. A
series of standard stars taken from the Perry, Olsen \& Crawford (1987)
and Olsen(1983) was also observed with the same procedure
on each night to transform the data into the standard system. The
seasonal transformation coefficients, except
for the 2002 season, are reported in Table 2. The indicated coefficients
are those in the following equations (Gronbech, Olsen \& Str\"omgren 1976).

\begin{eqnarray}
V = A + B~(b-y)_{st} + y_{inst}  \\
(b-y)_{st} = C + D~(b-y)_{inst}  \\
m_{1~st} = E + F m_{1~inst} + G ~(b-y)_{inst}  \\
c_{1~st} = H + I m_{1~inst} + J ~(b-y)_{inst} \\
\beta_{1~st} = K + L~\beta_{1~inst}
\end{eqnarray}

 For the 2002 season the transformation was made by filter instead of by
color indices. In this season the relations between the standard and the instrumental
values for each filter were the following:

\begin{equation}
V=y_{st}= a + b~ y_{inst}
\end{equation}
\begin{equation}
b_{st} =c + d~ b_{inst}
\end{equation}
\begin{equation}
v_{st}= e + f~ v_{inst}
\end{equation}
\begin{equation}
u_{st}= g + h~ u_{inst}
\end{equation}

\begin{figure}[!t]
\begin{center}
\includegraphics[width=8.cm,height=8.cm]{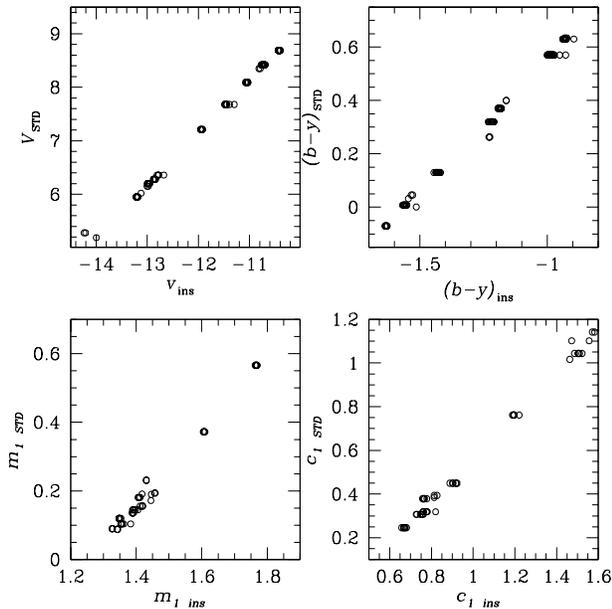}
\caption{Transformations from the instrumental to the standard $uvby$ system for the magnitude and colors. See eqs 1-5.}
    \label{fig1}
\end{center}
\end{figure}

\begin{figure}[t]
\includegraphics[width=8.cm,height=8.cm]{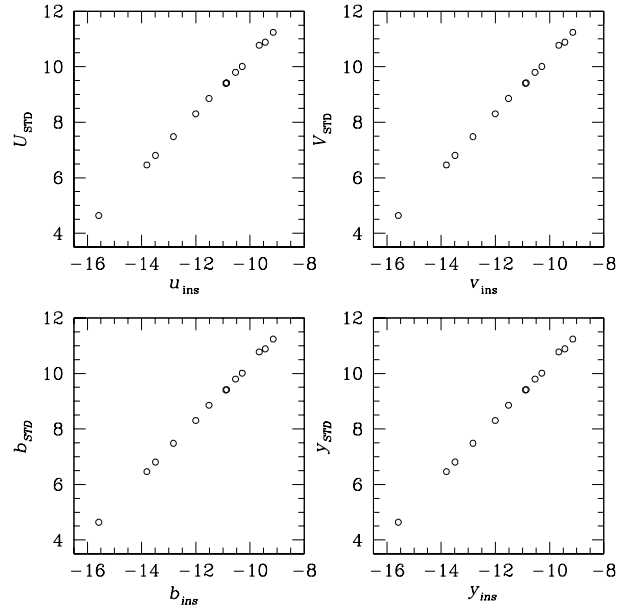}
\caption{Transformations from the instrumental $uvby$ magnitudes to the standard system.
See eqs. 6-9.}
    \label{fig2}
\end{figure}

\noindent
The coefficients are reported in the bottom part of Table 2.
The mean dispersions of the transformation equations are $0.028$,
$0.016$, $0.015$, $0.010$ mag in $u, v, b$ and $y$ respectively. It
was after this that the color indices were calculated. In Fig. \ref{fig1}
the instrumental $y$ magnitude and color transformations are shown.
In Fig. \ref{fig2} the relationships between the instrumental and
the standard magnitudes in the 2002 season are shown.

\begin{figure*}[ht]
\includegraphics[width=15.cm,height=12.cm]{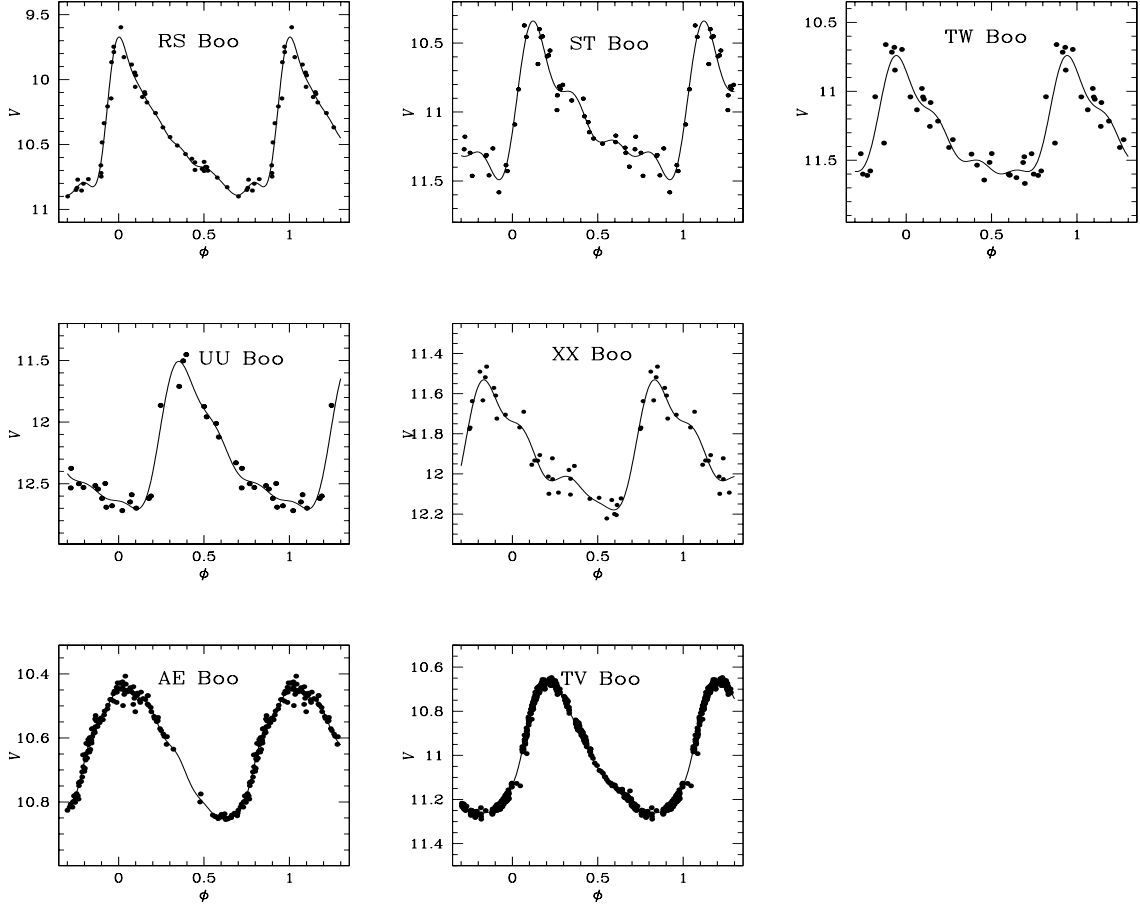}
\caption{V light curves of the sample stars.}
    \label{fig3}
\end{figure*}

The final $uvby-\beta$ photometry is available in electronic form from the Centre de Donn\'ees Astronomiques,
Strasbourg, France. The $V$ light curves are shown in Fig. \ref{fig3}.
In the following section a brief description of the light and color curves for each star is given.

\begin{table*}[!t]
\begin{center}
\setlength{\tabnotewidth}{\columnwidth}
  \tablecols{9}
\setlength{\tabcolsep}{1.0\tabcolsep}
 \caption{Transformation coefficients
of each season}
\label{COEFS}
  \begin{tabular}{lccccccrccccr}
\hline
season  &A & B      &C   &  D    &E &  F     &   G      &   H    &   I  &J &K &  L \\
\hline
Jun, 01  &18.076& $+0.020$ &1.354 & $0.991$&$-1.437$ & $1.012$ & $-0.002$&$-0.623$ & $1.001$ & $0.148$&$--$  &$--$ \\
Apr, 04  &18.839& $-0.010$ &1.539 & $0.993$&$-1.231$ & $0.949$ & $0.024 $&$-0.429$ & $0.988$ & $0.068$&$--   $  &$  --$ \\
May, 05  &19.000& $-0.007$ &1.531 & $0.983$&$-1.309$ & $1.068$ & $0.020 $&$-0.315$ & $1.031$ & $0.153$&$2.626$  &$-1.348$ \\
\hline
 \end{tabular}
\end{center}
\begin{center}
  \begin{tabular}{lcccccccc}
  & a         &  b     &  c     &   d      &   e    &   f   &  g & h  \\
\hline
Feb, 02 & 19.1565  &1.0054  & 20.9731   & 1.0236 & 21.0131 & 1.0101 & 20.5204 & 1.0025 \\
\hline
 \end{tabular}
\end{center}
\end{table*}

\section{Notes on individual stars}

Due to the fact that in most of the seasons the RR Lyrae stars were
not the primary target, the sampling of these stars was not
homogeneous. For example, AE Boo was observed for four
nights in 2001 for a relatively short time on each night, although the
phase coverage was almost complete. In the 2005 season, only a couple of
points per night were obtained for most RR Lyrae stars
although the time span of the observations covered seventeen nights.
For AE Boo around 180 points in V were gathered.

The other amply observed star was TV Boo which was measured in 2002 (250
data points), in 2004 (5 points) and in 2005 (13 points).
According to Wils et al. (2006)
TV Boo shows a Blazkho period of 10 d. Given the time span of our
observations on this star and the uncertainties in the V magnitude
(0.03 mag) we can neither verify nor contradict this assertion.
We call
attention to the fact that the time of maximum light of this star
does not coincide with a phase value of zero implying a secular
variation of the period. We did not pursue this possibility.

The last star which was
observed in more than one season was RS Boo. The data cover the
full cycle and the maximum light coincides with phase zero. Of the
remaining stars, ST Boo, TW Boo, UU Boo and XX Boo,
all observed in the 2005 season at the rate of a couple of points per night and
for seventeen nights, all
show a  maximum light shifted relative to phase
zero at least with the light elements reported in Table 3.
For each of these stars a sample of around thirty-five data points was gathered.

\begin{table*}[!t]

\small{
\begin{center}
\setlength{\tabnotewidth}{\columnwidth}
 \setlength{\tabcolsep}{1.0\tabcolsep}
 \caption{Ephemerides and Fourier coefficients for the sample stars.}
\label{COEF}
  \begin{tabular}{lcccccccccccc}
 \hline
ID & HJD & P & $A_0$  & $A_1$ & $A_2$ & $A_3$ & $A_4$ & $\phi _{21}$ & $\phi _{31}$ & $\phi _{41}$ & N & Type \\
& 2400000.0+ &  (days)   &$\sigma_{A_0}$ &$\sigma_{A_1}$ &$\sigma_{A_2}$ &$\sigma_{A_3}$ &$\sigma_{A_4}$ & $\sigma_{\phi _{21}}$& $\sigma_{\phi _{31}}$& $\sigma_{\phi _{41}}$& \\
\hline
RS Boo & 41770.4900 & 0.37733896 & 10.470 &0.447 &0.207 & 0.105& 0.084  &4.088 &1.881 &5.858 & 8 & ab\\
       &            &            &  0.014 &0.019 &0.018 & 0.020& 0.018 &0.129 &0.215 &0.287 &   &   \\
ST Boo & 19181.4860 & 0.62229069 & 11.044 &0.383 &0.194 & 0.133& 0.097 &3.979 &2.129 &6.112 & 4 & ab\\
       &            &            &  0.017 &0.023 &0.024 & 0.024& 0.024 &0.177 &0.268 &0.358 & & \\
TW Boo & 26891.2680 & 0.53227315 & 11.309 &0.359 &0.138 & 0.069& 0.058 &3.782 &1.827 &5.438 & 4 & ab\\
       &            &            &  0.027 &0.037 &0.039 & 0.035& 0.039 &0.359 &0.703 &0.772 &   &   \\
UU Boo & 36084.4100 & 0.45692050 & 12.251 &0.504 &0.203 &0.086 & 0.055 &3.905 &2.074 &6.078 & 4 & ab\\
       &            &            &  0.022 &0.030 &0.032 &0.035 & 0.035 &0.212 &0.431 &0.652 &   &   \\
XX Boo & 29366.6460 & 0.58140160 & 11.914 &0.257 &0.114 &0.051 & 0.039 & 4.087 & 2.726& 6.189 & 4 & ab\\
       &            &            &  0.012 &0.018 &0.015 &0.017 & 0.019 & 0.201 & 0.387& 0.476 &   &   \\
AE Boo & 30388.2500 & 0.31489240 & 10.646 &0.202 &0.027 &0.008 & 0.008 & 4.903& 3.722& 2.377& 6 & c \\
       &            &            &  0.004 &0.006 &0.006 &0.006 & 0.006 & 0.223& 0.730& 0.670&   &   \\
TV Boo & 24609.5150 & 0.31255936 & 11.020 &0.282 &0.078 &0.020 & 0.015 & 3.915& 1.666 & 6.043 & 7 & c \\
       &            &            &  0.001 &0.001 &0.001 &0.001 & 0.001 & 0.017& 0.051 & 0.074 &   &   \\
\hline
  \end{tabular}
\end{center}
}
\end{table*}

\section{Iron abundance estimation}

\subsection{{\rm [Fe/H]} from the Fourier light curve decomposition}

As a first approach, the V light curves were decomposed
in their harmonics and the Fourier coefficients used
to estimate the iron abundance via semiempirical calibrations.
In order to do that, the light curves were phased with periods and the epochs listed in
Table \ref{COEF}, which were adopted from Kholopov et al. (1985).

Given the ephemerides, the light curves in Fig. \ref{fig3} were fitted by the equation:

\begin{equation}
\label{fourier}
    m(t)=A_0 + \sum_{k=1}^N A_k \cos(\frac{2\pi}{P}k
    (t-E)+\phi_k)\nonumber
\end{equation}

where $k$ corresponds to the $k-th$ harmonic of amplitude $A_k$ and displacement $\phi _k$.

The Fourier coefficients, defined as:

\begin{eqnarray}
    \phi_{ij}& = & j\phi_i - i\phi_j \nonumber \\
       R_{ij}& = & A_i / A_j \nonumber
\end{eqnarray}

\noindent
have been calibrated in terms of physical parameters.
The Fourier coefficients corresponding to the solid curve in Fig. \ref{fig3}
and the number of harmonics used to produce the best possible fit,
are given in Table \ref{COEF}. The number of significant harmonics depends on the
dispersion of the light curve. Only significant harmonics were retained;  and their influence on [Fe/H] estimated from the Fourier decomposition will be discussed at the end of this section.

For the RRab stars, [Fe/H] was estimated using the calibration of
Jurcsik \& Kov\'acs (1996);

\begin{equation}
\label{JK}
    {\rm [Fe/H]}_{\rm J} = -5.038 ~-~ 5.394~P ~+~ 1.345~\phi^{(s)}_{31},
\end{equation}

\noindent
The standard deviation in the above equation is 0.14 dex. In eq. \ref{JK}, the phase $\phi^{(s)}_{31}$ is calculated from a sine series. To convert the cosine series based $\phi^{(c)}_{jk}$ into the sine series $\phi^{(s)}_{jk}$,
one can use ~$\phi^{(s)}_{jk} = \phi^{(c)}_{jk} - (j - k){\pi \over 2}$.
The metallicity [Fe/H]$_{\rm J}$ from eq. \ref{JK} can be converted to the
metallicity scale of Zinn \& West (1984) (ZW) via [Fe/H]$_{\rm J}$ = 1.43 [Fe/H]$_{\rm ZW}$ + 0.88 (Jurcsik 1995). The values of [Fe/H]$_{\rm ZW}$ are given in Table \ref{FEHDM}.

\begin{table}[!t]
\begin{center}
\setlength{\tabnotewidth}{\columnwidth}
  \tablecols{4}
 \setlength{\tabcolsep}{1.0\tabcolsep}
 \caption{[Fe/H]$_{ZW}$ for the sample of RR Lyraes. }
\label{FEHDM}
  \begin{tabular}{lccc}
 \hline
ID & [Fe/H]$_{ZW}$ & $D_m$ & Type \\
\hline
RS Boo& $-0.84$ & 4.1 & ab \\
ST Boo& $-1.53$ & 1.2 & ab \\
TW Boo& $-1.47$ & 1.9 & ab \\
UU Boo& $-0.95$ & 5.9 & ab \\
XX Boo& $-0.81$ & 5.1 & ab \\
AE Boo& $-1.30$ &  & c  \\
TV Boo& $-2.04$ &  & c \\
\hline
  \end{tabular}
\end{center}
\end{table}

Before applying eq.~\ref{JK} to the five RRab stars in our sample, we calculated the
$compatibility ~  condition ~ parameter$ $D_m$ which, according to
Jurcsik \& Kov\'acs (1996) and  Kov\'acs \& Kanbur (1998), should be smaller that 3.0.
The values of $D_m$ for the five RRab stars are given in Table \ref{FEHDM}.
It is worth commenting that the $D_m$ parameter calculated for the RRab stars does not seem
to correspond to the quality and/or density of the light curve. For instance, from
Fig. \ref{fig3} the curve of RS Boo is clearly the best and that of ST Boo is among the worst, however the corresponding $D_m$ values are 4.1 and 1.2 respectively, which is contrary to what would be expected. Thus we decided to relax this criterion a bit and apply the decomposition to the five RRab stars in our sample.
We shall discuss the uncertainties in [Fe/H] due to the number of harmonics used to fit the data later in this section.

For the RRc stars we have calculated [Fe/H] using the recent calibration of Morgan et al. (2007), which
provides [Fe/H] in the ZW scale;

\begin{eqnarray}
\label{MORG}
{\rm [Fe/H]}_{\rm {ZW}} = 52.466 P^2 - 30.075 P + 0.131 (\phi^{(c)}_{31})^2 \nonumber \\
+ 0.982 \phi^{(c)}_{31} -4.198 \phi^{(c)}_{31} P + 2.424
\end{eqnarray}

The results are listed in Table \ref{FEHDM}.

Eq. \ref{JK} is calibrated using the compilation of $\Delta S$ values of Suntzeff et al. (1994)
transformed into [Fe/H] and the spectroscopic values [Fe/H]$_K$ of Layden (1994). On the other hand,
eq. \ref{MORG} is calibrated using the [Fe/H] values of Zinn \& West (1984) and Zinn (1985) which are
the weighted average of iron abundances obtained from an assortment of methods, including the
 $\Delta S$ and the Q$_{39}$ indices. Therefore the values of [Fe/H] derived from the Fourier approach
for both RRab and RRc stars, are not completely independent from the values of [Fe/H] estimated from $\Delta S$, however
a discussion of $\Delta S$ and [Fe/H] values found in the literature for the stars in our sample is
of interest and we present it in the following section.

Perhaps the most relevant source of uncertainty in the physical parameters estimated from the Fourier decomposition approach is the quality and density of the light curve and hence its mathematical representation, alas, the number of
harmonics needed to fit the observed data. For the four RRab stars ST Boo, TV Boo,
UU Boo and XX Boo, whose light curves are scattered and/or not very dense, the number of significant harmonics 3 or 4 is rather undistinguishable.  It can be noted that the values of $T_{\rm eff}$ and $M_V$
vary 25-90~K and 0.01-0.07 mag. respectively, while the most striking variation
takes place in [Fe/H] where variations on average were $\pm 0.19$ dex. This value is marginally larger than the declared mean standard deviations of eqs.
\ref{JK} and \ref{MORG} (0.14 mag.) Thus we estimate the uncertainties in the values of [Fe/H]$_{ZW}$ of about $\pm 0.17$ dex. We have retained the cases with the smaller standard deviation from the fit and with exclusively significant harmonics in Table \ref{COEF}.

\subsection{{\rm [Fe/H]} from the $\Delta S$ parameter}

It is a common practice to estimate the value of [Fe/H] for RR Lyrae stars
using the $\Delta S$ metallicity parameter; Preston's (1959) method. In an extensive study of
the $\Delta S$ parameter on RR Lyrae variables in the Galactic halo, Suntzeff et al.
(1994) provided average values of  $\Delta
S$ with errors of $0.3$ units, which correspond to
$0.05$ dex in [Fe/H]. They reported $\Delta S$ values for three
stars of the present study: RS Boo, TW Boo and TV Bootes with values 1.36, 2.3 and 11.6. For RS Boo and TV Bootes
they give numerous estimations of $\Delta S$, but since they correspond to similar phases, we have taken the averages and marked them with asterisks in Table \ref{deltaS}. $\Delta S$ values for ST Boo, TW Boo and XX Boo
are found in the paper by Smith (1990), the average or individual values are given in Table \ref{deltaS}.
We note here the discrepant values for TW Boo from Suntzeff et al.
(1994) and Smith (1990) despite their corresponding to similar phases of 0.12 and 0.17 respectively. Also two discrepant values for XX Boo but for very different phases
are found in Smith (1990). A value
for TV Boo is given by Liu \& Janes (1990) as 12.1 which was measured near maximum light.
A summary of the $\Delta S$ values for our sample stars is given in Table \ref{deltaS}.

\begin{table*}[!t]
\small{
\begin{center}
\setlength{\tabnotewidth}{\columnwidth}
  \tablecols{7}
 \setlength{\tabcolsep}{1.0\tabcolsep}
 \caption{$\Delta S$ values for sample stars}
\label{deltaS}
  \begin{tabular}{lcccccr}
 \hline
ID &  Type & Spectra & $\Delta S$\tabnotemark{1} & $\Delta S$\tabnotemark{2} &$\Delta S$\tabnotemark{3} &  Px (mas)\\
\hline
RS Boo&   ab & A7-F5 & 1.36*$^m$ &          &  & $0.11\pm 1.40$\\
ST Boo&   ab & A7-F7 &      & 8.9* &  & $1.19\pm 1.61$\\
TW Boo&   ab & F0-F8 & 2.3  & 6.2,8.3$^m$ &  & $-0.28\pm 1.63$    \\
UU Boo&   ab &       &      &          &  &     \\
XX Boo&   ab &       &      & 7.0:,10.1: &  &     \\
AE Boo&   c  & F2    &      &          &  & $0.32\pm 2.00$ \\
TV Boo&   c  & A7-F2 & 11.6*$^m$ & & 12.1  &  $-0.07\pm 1.60$   \\
\hline
 \tabnotetext{1}{Suntzeff et al. (1994)}
 \tabnotetext{2}{Smith (1990)}
 \tabnotetext{3}{Liu and Janes (1990)}
 \tabnotetext{*}{average of multiple measurements }
 \tabnotetext{m}{values from near minimum light}
 \tabnotetext{:}{unknown phase}
\end{tabular}
\end{center}
 }
\end{table*}

\begin{table*}[!t]
\begin{center}
\setlength{\tabnotewidth}{\columnwidth}
  \tablecols{8}
 \setlength{\tabcolsep}{1.0\tabcolsep}
 \caption{Physical parameters from the calibrations for studied RR Lyrae stars. }
\label{FEMV}
  \begin{tabular}{lccccccccc}
 \hline
ID & [Fe/H]$_{ZW}$ & [Fe/H]$_{\Delta S}$ &[Fe/H]$_{K}$ & $T_{\rm eff}$ &$M_V$ & BC &$\log L/L_{\odot}$ & d (pc) &  N \\
\hline
RS Boo& $-0.84$ & $-0.81$ &$-0.32 \pm 0.09$ & 6905&0.79 & 0.00& 1.58 & 862  &   8 \\
ST Boo& $-1.53$ & $-1.82$ &$-1.86 \pm 0.14$ & 6454&0.48 & -0.05&1.73 & 1295 &   4 \\
TW Boo& $-1.47$ & $-0.90,-1.46,-1.74$ &$-1.41 \pm 0.09$ & 6581&0.58 & -0.05&1.69 & 1396 & 4 \\
UU Boo& $-0.95$ &         &$-1.92 \pm 0.20$ & 6818&0.56 & -0.01&1.68 & 2284 & 4 \\
XX Boo& $-0.81$ & $-1.56,-1.98$ & & 6675&0.62 & -0.03&1.67 & 1818 & 4 \\
AE Boo& $-1.30$ &         & & 7384&0.58 & 0.026&1.74 & 1032 & 6  \\
TV Boo& $-2.04$ & $-2.18$ & & 7199&0.59 & 0.026&1.75 & 1218 & 7 \\
\hline
  \end{tabular}
\end{center}
\end{table*}

While converting $\Delta S$ into [Fe/H] is a current practice, one should bear in mind that
the determination of $\Delta S$ is subject to numerous sources of uncertainty, as have been
discussed for example by Butler (1975), Smith (1986; 1990) and Suntzeff, et al. (1994),
e.g. $\Delta S$ estimated during the rising light can differ greatly from that
estimated during the declining brightness. The correction to minimum phase is done using
empirical curves on the $\Delta S-SpT(H)$-plane with considerable dispersion and personal
judgment in their final definition (Smith 1986; 1990), and the weighted mean finally reported value of
$\Delta S$ is often obtained from a few measurements randomly distributed along the pulsation
cycle of the star. Some of the above considerations may explain the large differences in the  existing
values of $\Delta S$ for a given star (see Table \ref {deltaS}).

Beside the above sources of uncertainty in $\Delta S$, we have to note that the transformation of
this observational parameter into [Fe/H] encountered serious problems: the
high dispersion abundance determinations in RR Lyrae are old (e.g. Butler 1975,
Butler \& Deeming 1979, Butler et al. 1982) while modern high signal-to-noise,
high dispersion digital data, analyzed with modern synthetic codes are non-existent. According to Manduca (1981), conversion of $\Delta S$ to [Fe/H] has calibration problems for the metal-rich and metal-poor domains and in fact his theoretical calibration is not linear. We have chosen however,
to use the more recent
empirical linear transformation of Jurcsik (1995); ${\rm [Fe/H]}_J = -0.190 \Delta S$  $-0.027$,
which is valid for ${\rm [Fe/H]}_J$ between 0.0 and $-2.3$ dex and for field and cluster
RR Lyraes, to transform the $\Delta S$ parameter
into iron abundances on the ZW scale.
We have first used the $\Delta S$ values in Table \ref{deltaS} and the formula of Jurcsik (1995), to calculate ${\rm [Fe/H]}_J$, and then brought this into the ZW scale as discussed in $\S$ 4.1; we have labeled these iron values as [Fe/H]$_{\Delta S}$ and they are listed in Table \ref{FEMV}.

For the sake of comparison, in columns 2-4 of  Table \ref{FEMV} we have listed
the values of the Fourier estimations of the metallicity, [Fe/H]$_{ZW}$, the values from
$\Delta S$, [Fe/H]$_{\Delta S}$ and the independent estimations from Layden (1994)
[Fe/H]$_{K}$. [Fe/H]$_{K}$ values are based on the strength of the Ca II K line and are in the ZW scale.

For RS Boo [Fe/H]$_{K}$ is rather discrepant but the agreement between [Fe/H]$_{ZW}$ and  [Fe/H]$_{\Delta S}$ is excellent.
For ST Boo and TW Boo the agreement between the three estimations is very good especially considering the implicit uncertainties in the values of  $\Delta S$.
Large discrepancies are noted for XX Boo and TV Boo. In the case of XX Boo one may argue that the coverage of the light curve can be considerably improved and hence the Fourier value could be expected to improve. However in the case of TV Boo, whose light curve is well defined and very dense, and whose
$\Delta S$ values are not scattered, we do not have an explanation handy.
Nevertheless, as an overall comparison of approaches to the determination of the iron abundance in RR Lyrae stars, and
considering the uncertainties involved in the estimation of the  $\Delta S$ parameter discussed above and the fact that in the Fourier solution the full shape of the light curve is included, we consider the Fourier decomposition a more solid approach.

\subsection{The effective temperature $T_{\rm eff}$}

The effective temperature can also be estimated from the Fourier coefficients. For the RRab stars we used the calibrations of Jurcsik (1998)

\begin{equation}
\label{teffab}
    log~T_{\rm eff}= 3.9291 ~-~ 0.1112~(V - K)_o ~-~ 0.0032~[Fe/H],
\end{equation}

\noindent
with

$$ (V - K)_o= 1.585 ~+~ 1.257~P ~-~ 0.273~A_1 ~-~ 0.234~\phi^{(s)}_{31} ~+~ $$
\begin{equation}
\label{color}
~~~~~~~ ~+~ 0.062~\phi^{(s)}_{41}.
\end{equation}

\noindent
Eq. \ref{teffab}
has a standard deviation of 0.0018 (Jurcsik 1998), but the accuracy of $log~T_{\rm eff}$ is mostly set by the color eq. \ref{color}. The error estimate on  $log~T_{\rm eff}$ is 0.003 (Jurcsik 1998).

For the RRc stars we have used the calibration of Simon \& Clement (1993);

\begin{equation}
    log T_{\rm eff} = 3.7746 ~-~ 0.1452~log~P ~+~ 0.0056~\phi^{(c)}_{31}.
\end{equation}

The values obtained from the above calibrations are reported in column 5
of Table \ref{FEMV}.

\section{Distance to the field RR Lyrae}

\subsection{$M_V$ from the Fourier parameters}

Absolute magnitudes of RR Lyrae can also be estimated from the Fourier parameters.
For the RRab one can use the calibration of  Kov\'acs \& Walker (2001);

\begin{equation}
\label{KWab}
M_V(K) = ~-1.876~log~P ~-1.158~A_1 ~+0.821~A_3 +K.
\end{equation}

\noindent
The standard deviations in the above equation is 0.04 mag.
The zero point of eq.~\ref{KWab}, K=0.43, has been calculated by Kinman (2002) using the prototype star RR Lyrae as calibrator, adopting the absolute magnitude $M_V= 0.61 \pm 0.10$ mag
for RR Lyrae, as derived by Benedict et al. (2002) using the star parallax measured by the HST.
Kinman (2002) finds his result to be consistent with the coefficients of the $M_V$-[Fe/H] relationship given by Chaboyer (1999) and Cacciari \& Clementini (2003). All these results are consistent with the distance modulus of the LMC of $18.5 \pm 0.1$ (Freedman et al. 2001; van den Marel et al. 2002; Clementini et al. 2003). Catelan \& Cort\'es (2008) have argued that the prototype RR Lyr has an overluminosity due to evolution of 0.064 $\pm$ 0.013 mag relative to HB RR Lyrae stars of similar metallicity. This would have to be taken into account if RR Lyr is used
as a calibrator of the constant $K$ in eq. \ref{JK}. Considering this,  Arellano Ferro et al. (2008b) have estimated a new value of $K = 0.487$.

For the sake of homogeneity and better comparison with previous results on luminosities of RR Lyrae stars (e.g. Arellano Ferro et al. 2008a,b), in the following we have adopted $K=0.43$.

For the RRc stars, the calibration of Kov\'acs (1998) was used;

\begin{equation}
\label{KWc}
M_V(K) = ~-0.961~P ~-0.044~\phi^{(s)}_{21} ~+4.447~A_4 + 1.261.
\end{equation}

The standard deviation in the above equation is 0.042 mag. In fact
we have propagated the errors, given in Table \ref{COEF}, in the
amplitudes $A_1$ and $A_3$ in eq. \ref{KWab} and in
$\phi^{(s)}_{21}$ and $\phi^{(s)}_{41}$ in eq. \ref{KWc} and found
that they produce
 uncertainties in $M_V$ $\sim 0.04$ mag.
Cacciari et al. (2005) have pointed out that in order for the eq. \ref{KWc} to surrender absolute magnitudes in agreement with the mean magnitude for the RR Lyrae stars in the LMC, $V_0 = 19.064 \pm 0.064$ (Clementini et al. 2003), the zero point
of the above equation should be decreased by 0.2$\pm$0.02 mag.
After this correction, we found the $M_V$ values for the RRc stars
AE Boo and TV Boo reported in Table \ref{FEMV}. These values of $M_V$ have been converted into
log $L/L_{\odot}$ (col. 5). The bolometric corrections for the average temperatures of RRab and RRc stars given in column 7 of Table \ref{FEMV}, were estimated from the
$T_{\rm eff}$-BC$_V$ from the models of Castelli (1999) as tabulated in Table 4 of Cacciari et al. (2005).

\begin{table*}[!t]
\begin{center}
\setlength{\tabnotewidth}{\columnwidth}
  \tablecols{8}
 \setlength{\tabcolsep}{1.0\tabcolsep}
 \caption{$\uv$ photoelectric photometry and reddening of $\delta$ Scuti stars in Bootes}
\label{delred}
  \begin{tabular}{lccccccc}
 \hline
star &  $V$   & $b-y$   & $m_1$  & $c_1$ & $\beta$ &  $E(b-y)$  & $uvby-\beta$  \\
 & &  &    &  &  &  & source  \\
\hline
$\iota$ Boo &  4.75 & 0.128 & 0.198 & 0.834 & 2.817 & 0.001 & 1\\
$\gamma$ Boo &  3.03 & 0.116 & 0.191 & 1.008 & 2.817 & 0.006& 1\\
$\kappa$$^2$Boo &  4.54 & 0.125 &  0.187 &  0.951&  2.806 & 0.001  & 1 \\
CN Boo  &  5.98 & 0.162 & 0.201 & 0.748 & 2.770  & 0.000 & 1\\
YZ Boo  & 10.57 & 0.151 & 0.178 & 0.868 & 2.768 & 0.000 & 1\\
YZ Boo  & 10.466 & 0.184 & 0.136  & 0.681 & 2.723 & 0.000 & 2 \\
\hline
 \tabnotetext{1}{Rodr\'{\i}guez et al. (1994)}
 \tabnotetext{2}{Pe\~na et al. (1999)}
 \end{tabular}
\end{center}
\end{table*}

\subsection{Reddening}

In order to determine distance and detailed variations of
the physical parameters along the cycle of pulsation of each star,
it is necessary to first estimate the reddening. For
field stars however, a proper determination of reddening is complex
and no direct method is known to us. In view of this, we have
determined the reddening of different objects in the same direction of the sky.
We have chosen five $\delta$ Scuti stars and the globular cluster NGC~5466 in Bootes
and M3 which is very near NGC~5466, as indicators of reddening in that direction of the sky.
Despite the large distance to M3 and NGC~5466 (10.4 and 15.9 kpc respectively),
$E(b-y)\sim 0.0$ as estimated from the $E(B-V)$ values listed by Harris (1996).
In the case of the $\delta$ Scuti stars we have used the expressions derived by
Crawford (1975, 1979) for F- and A-type stars respectively, with the zero point correction
suggested by Nissen (1988), to estimate the intrinsic color $(b-y)_o$ for
stars near the main sequence. Two sources of $uvby-\beta$ were considered;
Rodr\1guez et al. (1994) and Pe\~na et al. (1999). These $\delta$ Scuti stars,
their magnitude-weighted mean colors and $E(b-y)$ are listed in Table \ref{delred}.
>From these results it seems reasonable to conclude that the
reddening for the sample RR Lyrae stars in Bootes is negligible and we shall assume a value of
zero for all the stars in our sample.

\begin{figure}[t]
\includegraphics[width=8.cm,height=8.cm]{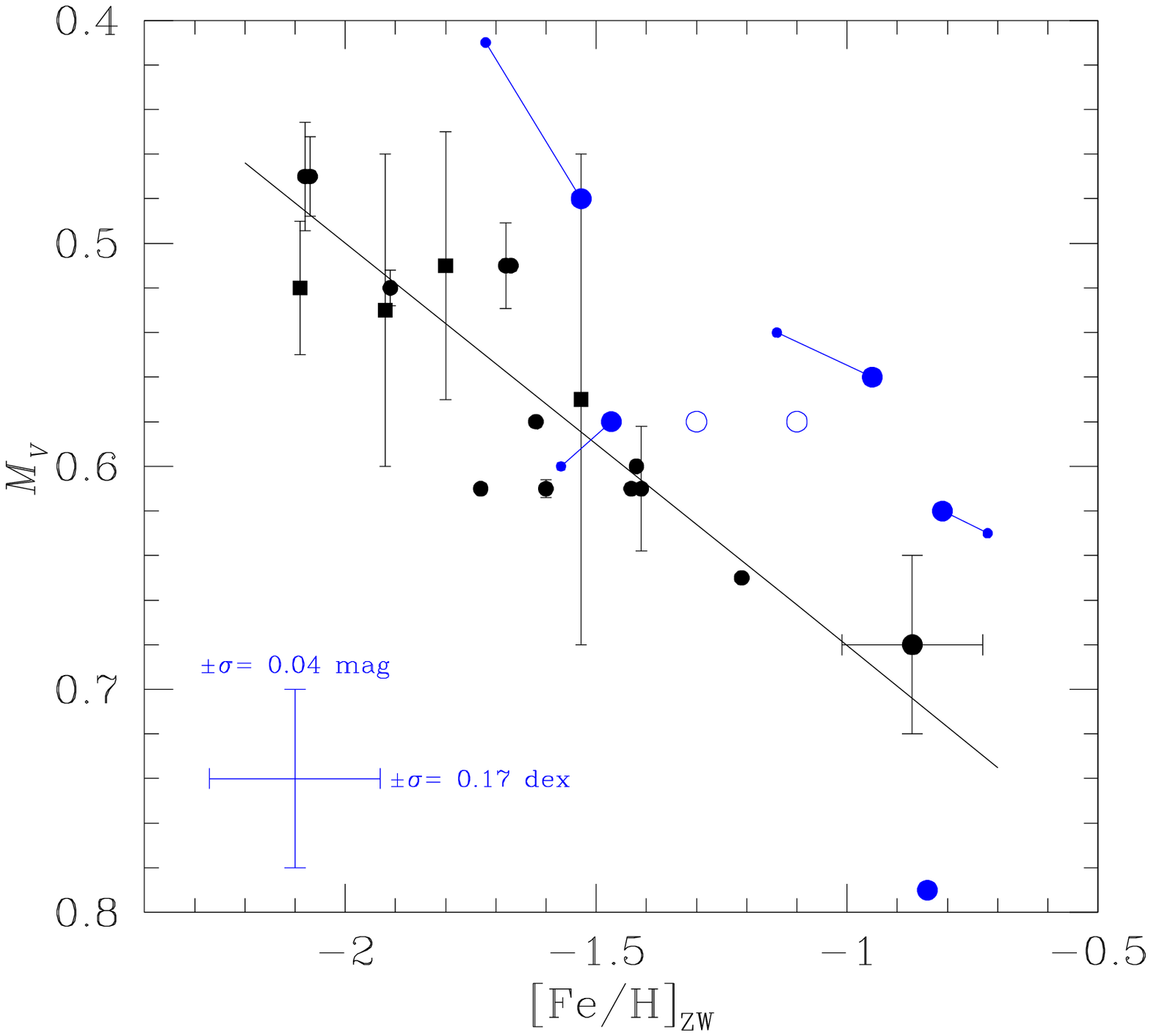}
\caption{$M_v- {\rm [Fe/H]}$ relation derived from cluster RR Lyrae stars (Arellano Ferro et al. 2008b). Blue points are the five RRab stars in our sample. When available, two positions are given; the larger points are the finally adopted solution while the smaller points correspond to similar solutions with smaller number of harmonics. Open blue circles correspond to the two RRc stars in our sample.
The uncertainties in the values of the blue symbols are discussed in the text and indicated by the
blue error bars.}
    \label{fig4}
\end{figure}

\subsection{Distances}

The implied distances given by the Fourier absolute magnitudes, and the reddenings and bolometric
corrections discussed in $\S$ 5.1 and 5.2 are listed in column 9 of Table \ref{FEMV}. These
distances can be compared with estimations from the parallaxes,
for those stars which have parallaxes determined by the new reductions of the Hipparcos data by
van Leeuwen (2007). For the stars
RS Boo, ST Boo, TW Boo, AE Boo and TV Boo the parallaxes and their errors are listed in Table \ref{deltaS}.
It should be noted that their errors are very large and hence the parallaxes for these stars very uncertain.
Except for ST Boo, those numerical values lead to distances much larger than the values derived from
the Fourier approach in Table \ref{FEMV}, and if these distances are used to calculate the
corresponding absolute magnitude they lead to absurd results.

\begin{figure*}[t]
\includegraphics[width=16.cm,height=16.cm]{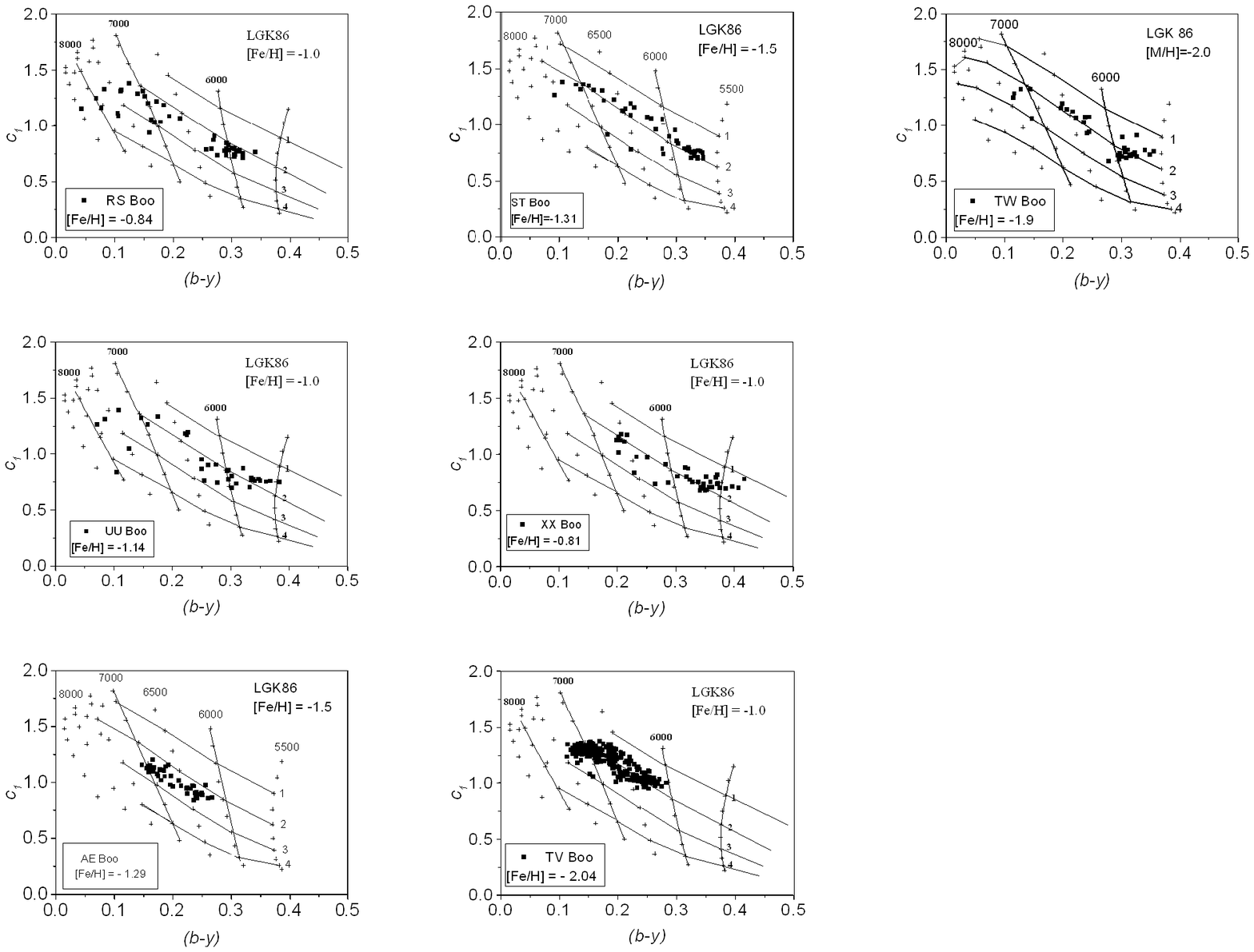}
\caption{Model grids on the $(b-y)_0 - c_0$ plane along the stellar variations through the cycle.}
    \label{fig5}
\end{figure*}

\section{$M_v$ from the Str\"omgren $\c1$ index}

The Str\"omgren $c_1$ index is a gravity (hence luminosity)
sensitive parameter for late-A to F type stars (Str\"omgren 1966), and therefore
it is thought to be useful in
determining the absolute magnitude $M_V$ for RR Lyrae stars if properly
calibrated. In fact, in a recent paper Cort\'es \& Catelan (2008) have offered such calibration for RR Lyrae stars.
These authors have employed the
pseudocolour $c_0 \equiv (u-v)_0 -(v-b)_0$, the fundamental period $P$
and the metallicity $Z$ for a large number of synthetic RR Lyrae stars
to calibrate $M_y$ and the colors $b-y$, $v-u$ and $u-v$ (their eq. 1).

\begin{table*}
\small{
\begin{center}
\setlength{\tabnotewidth}{\columnwidth}
  \tablecols{7}
 \setlength{\tabcolsep}{1.0\tabcolsep}
 \caption{Comparison of $M_V(F)$ obtained from the Fourier approach with two
theoretical calibrations from Cort\'es \& Catelan (2008).}
\label{MVC0}
  \begin{tabular}{lcccccc}
 \hline
 Star  &  [Fe/H]$_{ZW}$ & $(c_1)$ & $M_y(c1)$\tabnotemark{1} & $M_y(Z)$\tabnotemark{2} & $M_V(F)$\tabnotemark{3}& $M_V(F)$-$M_y(c1)$ \\
 \hline
RS Boo & $-0.84$ & ~~0.928 &~~0.884 & ~~0.765 & 0.79 & $-0.093$\\
 & & $\pm 0.012$ & $\pm 0.016$ & $\pm 0.063$ & & \\
ST Boo & $-1.53$ & ~~0.922 &~~0.319 & ~~0.526 & 0.48 & $+0.163$\\
 & & $\pm 0.010$ & $\pm 0.013$ & $\pm 0.030$ & & \\
TW Boo & $-1.47$ & ~~0.893 &~~0.537 & ~~0.577 & 0.58 & $+0.045$\\
 & & $\pm 0.017$ & $\pm 0.022$ & $\pm 0.039$ & & \\
UU Boo & $-0.95$ & ~~0.941 &~~0.659 & ~~0.665 & 0.56 & $-0.098$\\
 & & $\pm 0.017$ & $\pm 0.023$ & $\pm 0.051$ & & \\
XX Boo & $-0.81$ & ~~0.814 &~~0.678 & ~~0.776 & 0.62 & $-0.057$\\
 & & $\pm 0.011$ & $\pm 0.018$  & $\pm 0.063$ & & \\
AE Boo & $-1.30$ & ~~1.002 &~~0.880 & ~~0.619 & 0.58 & $-0.298$\\
 & & $\pm 0.003$ & $\pm 0.004$ & $\pm 0.044$  & & \\
TV Boo & $-2.04$ & ~~1.133 &~~0.625 & ~~0.481 & 0.59 & $-0.034$\\
 & & $\pm 0.006$ & $\pm 0.006$ &  $\pm 0.018$ & & \\
\hline
 \tabnotetext{1}{from $C_0, Z ~and~ P$ through eq. 1 of Cort\'es \& Catelan (2008).}
 \tabnotetext{2}{from $Z$ through eq. 5 of Cort\'es \& Catelan (2008).}
 \tabnotetext{3}{from the Fourier decomposition approach in this work, Table \ref {FEMV}.}
 \end{tabular}
\end{center}
}
\end{table*}

We have fitted the $c_1$ curves with a curve of the form of eq.
\ref{fourier} to estimate the magnitude-weighted means $(c_1)$ given
in column 3 of Table \ref{MVC0}. These magnitude-weighted means
differ from the intensity weighted means $<c_1>$ by about 0.01 mag.
(Catelan \& Cort\'es, 2008), which are smaller than the uncertainties for
$c_1$ of our observations as derived from the standard stars (see
Table 3 in Pe\~na et al., 2007). Since $E(b-y) = 0.$ for the sample
stars (sec. 5.2) we have taken $c_0 = c_1$. For the conversion of
[Fe/H] to Z we have used log Z = [M/H] - 1.765; M/H = [Fe/H] + log
(0.638 $f$ + 0.362) and $f = 10^{[\alpha /Fe]}$ (Salaris et al.
1993). We adopted $\alpha = 0.31$ as an appropriate value for halo
population stars. The predicted values of the absolute magnitude
from $c_1$, $M_y(c1)$, are given in column 4 of Table \ref{MVC0}.

In their equation 5 Cort\'es \& Catelan (2008) have calculated a
quadratic calibration of the $M_y$ - log Z relationship which can be
used to calculate $M_y(Z)$ given in column 5 of Table \ref{MVC0}. As
argued by Cort\'es \& Catelan (2008) $M_y(c1)$ refers to the
magnitude of an individual star whereas $M_y(Z)$ is the absolute
magnitude of a star of similar metallicity. Thus we shall compare
the Fourier approach results of each star, $M_V(F)$, with their
corresponding $M_y(c1)$. This comparison is made in column 7. The
uncertainties of $M_y(c1)$  are the resutl of propagating the
uncertainties in $<c_1>$  throught the equation 1 of Cort\'es \&
Catelan (2008). The uncertainties $M_y(Z)$ are estimated by
propagating the uncertinty in $\sigma _{[Fe/H]} = 0.17$ dex ($\S$
4.1). One can see that the dispersion in the calibration of equation
1 in Cort\'es \& Catelan (2008) is $\leq$ 0.01 mag (their Fig. 3)
and we recall that the uncertainties in $M_V(F)$ is $\pm 0.04$ mag
($\S$ 5.1). Therefore the differences in column 7 seem to be a bit
on the large side. It should be noted, however, that no systematics
can be seen and that if the $M_y(c1)$ values are plotted in Fig.
\ref{fig4} the dispersion of the field stars about the cluster $M_V-
{\rm [Fe/H]}$ relationship becomes very large. Also it can be seen
that the values of $M_y(c1)$ for the two RRc stars, AE Boo and TV
Boo are the most discordant despite of having the most densely
covered light curves and hence the smallest uncertainties.

\section{$M_V - {\rm [Fe/H]}$ relation}

A recent linear version of the $M_V- {\rm [Fe/H]}$ relationship for RR Lyrae stars has been calculated by
Arellano Ferro et al. (2008a) based on the light curve decomposition technique of RR Lyrae stars in a
group of globular clusters with a large range of metallicities. This relationship, reproduced in Fig \ref{fig4}, has been amply discussed by Arellano Ferro et al. (2008b) who found it to be consistent with independent
empirical linear versions and with theoretical non-linear versions after evolution from the Red Giant Branch is taken into account.
To check the consistency of our present results for the field stars in Bootes with those of RR Lyrae stars in globular clusters, we have plotted  in Fig \ref{fig4} the seven stars in our sample using the
Fourier [Fe/H]$_{ZW}$ and $M_V$ reported in Table \ref{FEMV}. It can be seen that with some larger scatter the Bootes stars distribution, given the uncertainties, follow the trend of the globular cluster RR Lyrae. The error bars correspond to the uncertainties in the Fourier calibrations of [Fe/H]$_{ZW}$ and $M_V$, i.e., those of eqs. \ref {JK}, \ref {MORG}, \ref {KWab} and \ref {KWc}.

\begin{table*}
\small{
\begin{center}
\setlength{\tabnotewidth}{\columnwidth}
  \tablecols{14}
 \setlength{\tabcolsep}{1.0\tabcolsep}
 \caption{$T_{\rm eff}$ and $\log~g$ variation ranges from the theoretical grids of LGK86}
\label{LGK86}
  \begin{tabular}{lccccccccc}
 \hline
star & [Fe/H] & $T_{\rm eff}$&  $T_{\rm eff}$ &  $T_{\rm eff}$ &
$\Delta T_{{\rm eff}}$ &  $\log g$ &
$\log g$ & $\Delta\log g$ &  $\Delta V$  \\
   & adopted & Fourier & min &  max & $K$ & min & max & dex &  mag \\
\hline
RS Boo & -1.0 & 6569 & 5700 & 8000 & 2300 & 2.2 & 3.5 & 1.2 &  0.82 \\
ST Boo & -1.5 & 6141 & 5700 & 7500 & 1800 & 1.5 & 3.0 & 1.5 &  0.73 \\
TW Boo & -2.0 & 6262 & 5500 & 7000 & 1500 & 1.5 & 3.0 & 1.5 &  0.85 \\
UU Boo & -1.0 & 6486 & 5500 & 8000 & 2500 & 1.5 & 3.0 & 1.5 &  0.63 \\
XX Boo & -1.0 & 6350 & 5000 & 6800 & 1300 & 1.5 & 3.0 & 1.5 &  0.77 \\
AE Boo & -1.5 & 7384 & 6100 & 7200 & 1100 & 1.9 & 2.8 & 0.9 &  0.48 \\
TV Boo & -1.0 & 7199 & 6000 & 7500 & 1500 & 1.3 & 2.5 & 1.2 &  0.79 \\
\hline
 \end{tabular}
\end{center}
}
\end{table*}

\section{Determination of physical parameters along the pulsational cycle}
Given the
simultaneity in the acquisition of the data in the different color
indices, once the reddening
has been inferred, it is possible to determine the variation of the
physical parameters of the star along the cycle.
This can be accomplished with the models developed
particularly for $\uv$ photometry by Lester, Gray \& Kurucz
(1986) (LGK86). The models have been built taking into account that the
$\uv$ system is well designed to measure key spectral signatures
that can be used to determine basic stellar parameters. The
theoretical calibrations have the advantage of relating the
photometric indices to the effective temperature, surface gravity,
and metallicity. LGK86 provide grids, on the plane $(b-y) - c_1$,
of constant $T_{\rm eff}$ and log g, for a large range of [Fe/H] values.
Based on the Fourier value [Fe/H]$_{ZW}$, a model with the nearest [Fe/H] value
was chosen for each star.
In Fig. \ref{fig5} the cycle variation of each star on its corresponding grid is illustrated.
This allows an estimation of the $T_{\rm eff}$ and log g variation ranges during the pulsation
cycle and a comparison with the estimated temperature from the Fourier approach.

It is interesting to note that the studied stars have different
effective temperatures and surface gravity limits, as well as
different ranges which cannot be determined with detail with only
the Fourier techniques. These results have been summarized in Table \ref{LGK86}
in which we have also included those determined through the
empirical calibrations from the Fourier coefficients. As can be
seen, both methods give analogous results. In Table \ref{LGK86}
the variation ranges in V, $T_{\rm eff}$ and log g are also indicated.

\section{Conclusions}

>From data acquired in several photometric campaigns we have obtained
extended $uvby-\beta$ photometry over a relatively large time span
for two stars and data that adequately cover the cycle of pulsation
for all of them. The $V$ light curves have been Fourier decomposed
and the corresponding Fourier parameters from their harmonics were
used to calculate the iron abundance and luminosity of each star.
The reddening was estimated by considering different objects in
the same direction.  The unreddened Str\"omgren indices $c_0$ and
$(b-y)_0$ served to determine the variation along the cycle of
the effective temperature and surface gravity.

 The iron abundance [Fe/H]$_{ZW}$ was calculated
first from the Fourier decomposition of the light curve and the
calibration proposed by Kov\'acs and co-workers and described with
detail in $\S$ 4.1. We also utilized the $\Delta S$
parameter to estimate the metallicity. In $\S$ 4.2 we amply discuss
  the uncertainties and the limitation of this technique. We
  conclude that the Fourier decomposition approach gives more
reliable results.

Since RR Lyrae stars are distance indicators, the individual estimations
of the absolute magnitude for field stars from independent methods is of interest.
The absolute magnitude, $M_V (F)$, predicted from Fourier decomposition ($\S$ 5.1),
is reported with other two determinations.
Once the iron abundance and the reddening were determined, an
independent estimate of $M_V (c_1)$ can be made from the $c_1$ index and the pseudocolor
$C_0$ (Cort\'es \& Catelan 2008). This is described in $\S$ 6. Also
in that section
we calculated  $M_V (Z)$ from the metallicity alone making use of a theoretical
quadratic  $M_V - Z$ relationship offered by Cort\'es \& Catelan (2008).
All these
results were compiled in Table \ref{MVC0}. It was pointed out that, although in some individual cases the differences between $M_V (F)$ and for example $M_V (c_1)$ were larger than the
expected from the uncertainties of the methods involved,
no systematic trends could  be seen and in some cases the agreement is fairly good.
The agreement does not seem to be related to the quality or the density of the
$V$ light curve, since the disagreement is largest for the two RRc stars which also
have the best light curves in our sample.

The $M_V- {\rm [Fe/H]}$ relationship for RR Lyrae stars has been amply discussed in the
literature. A recent version of it calculated exclusively from the Fourier
decomposition approach of RRab and RRc
stars in globluar clusters (Arellano Ferro et al. 2008b) was used to confront our
present results for the field stars (Fig \ref{fig4}).
It can be seen that with some larger
scatter the distribution of the Bootes stars follow the same trend of the
globular cluster RR Lyrae stars and it was noted that if the alternative results
for the absolute magnitude,  $M_V (c_1)$ or $M_V (Z)$ were used the quality of the comparison would remain.

With $M_V (F)$ and the reddening we have reported the resulting distances for the
sample stars (Table \ref{MVC0}).
It would be desirable to compare these distances with the derived distances from
an independent technique. Unfortunately, those determined from the
new reductions of the Hipparcos catalogue (van Leeuwen 2007) have such large errors that the comparison is impossible.

\vspace{0.5cm}

Acknowledgements. We would like to thank the following people for
their assistance: the staff of the OAN, M. S\'anchez and V. Alonso
at the telescope during some of the 2005 observations and A. Pani
for those of 2002, and L. Parrao for the reduction of the 2002
season. We are grateful to an anonymous referee for useful
corrections and suggestions. This paper was partially supported by
PAPIIT-UNAM IN108106 and IN114309. J. Miller and J. Orta did the
proofreading and typing, respectively. This article has made use of
the SIMBAD database operated at CDS, Strasbourg, France and ADS,
NASA Astrophysics Data Systems hosted by Harvard- Smithsonian Center
for Astrophysics.

\end{document}